\documentclass[twocolumn]{aastex62}
\usepackage{graphicx}
\usepackage{amssymb}
\usepackage{amsmath}
\usepackage{color}
\usepackage{hyperref}

\newcommand{\asat}{\emph{AstroSat}}

\newcommand{\sw}[1]{\texttt{#1}}

\newcommand{\tsearch}{\ensuremath{t_\mathrm{search}}}
\newcommand{\tbin}{\ensuremath{t_\mathrm{bin}}}

\newcommand{\qfar}{\ensuremath{F_\mathrm{quad}}}

\accepted{October 28, 2019}
\submitjournal{ApJ}

\shorttitle{CZTI upper limits on X-rays from FRBs}
\shortauthors{Anumarlapudi et al.}

\begin{document}

\title{Prompt X-ray emission from Fast Radio Bursts --- Upper limits with AstroSat}

\author[0000-0002-8935-9882]{Akash Anumarlapudi}
\affiliation{Indian Institute of Technology Bombay\\
Powai, Mumbai 400076, India}

\author[0000-0002-6112-7609]{Varun Bhalerao}
\affiliation{Indian Institute of Technology Bombay\\
Powai, Mumbai 400076, India}

\author[0000-0003-2548-2926]{Shriharsh P. Tendulkar}
\affiliation{Department of Physics and McGill Space Institute, McGill University,\\ 3600 University St, Montr\'eal, QC, Canada H3A 2T8}

\author[0000-0003-0477-7645]{A. Balasubramanian}
\affiliation{Indian Institute of Science Education and Research Pune\\ 
Pashan, Pune 411008, India}
\affiliation{Indian Institute of Technology Bombay\\
Powai, Mumbai 400076, India}
% \nocollaboration

\begin{abstract}
Fast Radio Bursts (FRBs) are short lived ($\sim$ msec), energetic transients (having a peak flux density of $\sim$ Jy) with no known prompt emission in other energy bands. We present results of a search for prompt X-ray emissions from 41 FRBs using the Cadmium Zinc Telluride Imager (CZTI) on \asat\ which continuously monitors $\sim70\%$ of the sky. Our searches on various timescales in the 20--200~keV range, did not yield any counterparts in this hard X-ray band. We calculate upper limits on hard X-ray flux, in the same energy range and convert them to upper bounds for $\eta$: the ratio X-ray to radio fluence of FRBs. We find $\eta \leq 10^{8-10}$ for hard X-ray emission. Our results will help constrain the theoretical models of FRBs as the models become more quantitative and nearer, brighter FRBs are discovered.
%\edit1{However, these results are unable to constrain theoretical models since these models are not very quantitative about the bounds on emissions in non-radio bands. (This can be more subtle and to the point).}
\end{abstract}


\section{Introduction} \label{sec:intro}
Fast Radio Bursts (FRBs) are bright ($\sim$\,Jy), spatially-unresolved and short ($\sim$ms duration) transients in the radio regime (frequency range of 400 MHz to 8 GHz). These are characterized by their high observed dispersion measures (DMs) --- often an order of magnitude higher than the total Galactic electron column density along the line of sight \citep{2017ApJ...835...29Y,Cordes:2002wz} --- indicating that the progenitor is extragalactic. The millisecond duration of the pulse constrains the emission region of the source to $r\lesssim c t_{\mathrm{pulse}} \sim$ 300 km, not considering any relativistic effects in the source frame. 
%The component frequencies follow a quadratic frequency dependent time delay over the width of the burst given as 
%\begin{equation}
%\label{eqn:dispersion_relation}
%\left(\frac{t_\mathrm{DM}}{\mathrm{ms}}\right) = 4.1488 \left( \frac{\mathrm{DM}}{\mathrm{pc~cm}^{-3}} \right) \left( \frac{\nu}{\mathrm{GHz}} \right)^{-2}
%\end{equation}

A total of 85 FRB detections have been publicly reported till September 2019 \citep{2016PASA...33...45P}. Of these, 11 FRBs \citep{RepFRB121102,RepFRBCHIME1,RepFRBCHIME8, RepFRB171019} are found to be repeating, but no periodicity or pattern has been found in its repetition \citep{2016Natur.531..202S, 2016ApJ...833..177S}. Unlike other FRBs, FRB 121102 has been localized, to milli-arcsecond precision, to a dense star forming region of a low-metallicity dwarf galaxy with redshift $z=0.193$ co-located within a projected transverse distance of 40 pc to an unresolved radio source~\citep{2017Natur.541...58C, 2017ApJ...834L...7T, 2017ApJ...834L...8M}. This localization and redshift measurement has led to a detailed study of its energetics, host environment \citep{2017ApJ...843L...8B, 2017ApJ...844...95K} and possible links to long Gamma-ray bursts (GRBs) and hydrogen-poor superluminous supernovae \citep{2017ApJ...841...14M, 2018arXiv180605690M, 2018arXiv180809969M}.

%Models proposed
Until now, no clear physical picture of either the mechanism for an FRB emission or the progenitor has emerged. A wide range of models have been proposed, many of which invoke neutron stars and strong magnetic fields to explain the short duration and high brightness temperatures of FRBs. The astrophysical scenarios hypothesized for the origin of FRBs include Crab-like giant pulses from neutron stars \citep{Cordes:2015fua,2017MNRAS.469L..39K}, magnetar giant flares \citep{2002ApJ...580L..65L, 2013arXiv1307.4924P, 2014ApJ...797...70K, 2015ApJ...807..179P}, binary neutron star mergers \citep{2013PASJ...65L..12T, 2018arXiv180804822P}, collisions between asteroids and neutron stars \citep{2015ApJ...809...24G}, the collapse of a neutron star into a black hole \citep{2014A&A...562A.137F}. There are also non-neutron star models that hypothesize FRBs arising from processes such as Dicke's superradiance \citep{2018MNRAS.475..514H}, axion decay in a strong magnetic field \citep{2018arXiv180602352V}and compact explosions of macroscopic magnetic dipoles \citep{2017ApJ...844..162T}. See \citet{Platts:2018hiy}\footnote{Refer \url{https://frbtheorycat.org/} for a complete summary of proposed theoretical models.}; %\rough{and Weltman et al. \emph{in preparation}} 
\citep{2018arXiv180603628P} for a recent review of FRB observations and models.

The presence or absence of a prompt emission corresponding to FRBs in different wavebands can constrain the emission mechanisms. In models invoking curvature radiation, photons are emitted along the direction of electron motion and the scope for inverse Compton scattering to higher wavebands is small \citep{2017MNRAS.468.2726K, 2018A&A...613A..61G}. If at all, such models predict possible prompt counterparts to FRBs in the THz --- optical/infrared regime, but not at X-ray energies. Synchrotron emission models allow for possible inverse Compton upscattering of radio photons to X-ray energies, suggesting prompt X-ray/$\gamma$-ray counterparts for FRBs. Similarly, astrophysical scenarios such as binary neutron star mergers may also lead to the ejection of a GRB jet, which if aligned along our line of sight, will produce a short $\gamma$-ray burst. Radio observations, combined with X-ray, gamma-ray and gravitational wave observations will allow us to constrain the emission mechanisms of FRBs as well as to possibly discover the astrophysical scenarios that lead to them. \citet{Totani:2013lia} states that in particular, if some FRBs are linked to binary neutron star mergers and short GRBs, this may allow us to increase the detection horizon of LIGO and other gravitational wave observatories.

To date high-energy limits on FRB counterparts have been unconstraining due to the relative insensitivity of X-ray/$\gamma$-ray telescopes. \citet{tkp16} set limits on the fluence ratio in the $\gamma$-ray to radio bands of  $F_\mathrm{\gamma}/F_\mathrm{1.4\,GHz} \lesssim 10^{-7} - 10^{-9}\,\mathrm{erg\,cm^{-2}\,Jy^{-1}\,ms^{-1}}$ corresponding to $F_\mathrm{\gamma}/F_\mathrm{1.4\,GHz} \lesssim 10^{8}-10^{10}$ for a bandwidth of 1\,GHz\footnote{In the following discussion, the radio/X-ray/$\gamma$-ray limits that are stated arise from different wavebands and strict comparison would require conversion with a knowledge of the spectral energy distribution. Given the order-of-magnitude state of observational knowledge and theoretical constraints, we are using these without conversion.}. For most FRBs, these limits are inconsistent with the observational limits for radio emission from the 2007 giant flare of SGR\,1806$-$20 \citep{tkp16}.  For FRB 121102, \citet{2017ApJ...846...80S} set deep limits of $F_\mathrm{\gamma}/F_\mathrm{1.4\,GHz} < 10^{6}-10^{8}$ using simultaneous radio and X-ray observations. We caution that observed values (or limits) of the $\gamma$-ray to radio fluence may significantly differ from intrinsic (source-frame) values due to beaming effects.

There have been a number of efforts at low to intermediate radio frequencies ($\sim$ GHz) searching for prompt counterparts to GRBs, also without significant success. \citet{bannister2012} searched for short, dispersed radio transients from nine GRBs and detected a few $>6-\sigma$ candidates from two of them with delays of a few hundred seconds compared to the gamma-ray emission. However, the possibility of these being radio frequency interference could not be ruled out. \citet{palaniswamy2014} searched for radio transients within 140\,s of the occurrence of five GRBs but did not detect any significant candidate counterparts. The response time is limited by how fast a large radio dish can slew. In the future, these efforts may be improved by the chances of detecting GRBs in the fields of views of wide field radio telescopes or using software beamforming radio telescopes.

From synchrotron emission, typical expected X-ray to radio ratios are  $\sim10^{4}$ \citep{2002ApJ...580L..65L} to $10^{6}$ \citep{2014MNRAS.442L...9L}. Pulsars have typical $\gamma$-ray to radio ratios of $10^{4}-10^{8}$. Other theories like \citet{Katz:2015ltv}, \citet{Falcke:2013xpa} also predict prompt X-ray/$\gamma$-ray emission, though the emission ratios are not well-quantified. These constraints still allow for the possibility of FRBs originating from pulsars, magnetars or other astrophysical scenarios. However, as we discover greater numbers of FRBs, it is important to have a framework to easily search for counterparts and set limits. After we amass a large population of FRBs, especially radio-bright ones, with X-ray non-detections, we can use them to constrain FRB models.

The Cadmium Zinc Telluride Imager \citep[CZTI;][]{czti} on board \asat~\citep{astrosat} is a coded aperture mask instrument with a 4\fdg6 $\times$ 4\fdg6 imaging field of view in the hard X-ray band from 20~keV to 200~keV. The instrument structure is designed to be nearly transparent to photons with energies $\gtrsim 100~$keV, making CZTI sensitive to transients from the entire sky, barring $\sim30\%$ occulted by the Earth. The four identical, independent quadrants of CZTI allow for rejection of spurious signals to a high degree. CZTI has detected over two hundred GRBs\footnote{CZTI GRB detections are reported regularly on the payload site at \url{http://astrosat.iucaa.in/czti/?q=grb}.}. For transients with a clear detection, CZTI data can yield spectra~\citep{rch+16}, polarisation~\citep{vcr+15,cva+17} and localisation~\citep{2016ApJ...833...86R,2017ApJ...845..152B}. 

Here we present a framework for burst searches, the hard X-ray limits on known FRBs and a plan for a near-automated future pipeline.

\section{Data Analysis}
The standard procedure for CZTI transient searches differs from the usual analysis for steady sources. We outline the transient search method (used, for instance, in \citealt{2018ATel11417....1A, 2018ATel11413....1A}) in this section. In \S\ref{sec:results} we expand the sample with searches in archival data.
%\outline{CZTI data --- reduction --- qualitative search (TE plots) --- quantitative cutoffs --- detrending}

%

\subsection{Data reduction}\label{subsec:reduction}
CZTI data consists of time-tagged event mode data for individual photons in topocentric frame. \asat~
is in Low Earth Orbit (LEO) and since the time difference between the ground observatories and \asat~ is much smaller compared to search window(\tsearch), we do not change the frame of reference. The standard CZTI pipeline\footnote{The CZTI pipeline software is available at \url{http://astrosat-ssc.iucaa.in/?q=data_and_analysis}.} flags any short duration count rate spikes as noise and removes them from the data. Since this would remove any X-ray photons from FRBs (or Gamma Ray Bursts), we start with unprocessed Level-1 data. The first step is to run \sw{cztbunchclean} to remove photons created by particle interactions with the satellite. Good Time Intervals (GTIs) are estimated by \sw{cztgtigen} and data is selected from these GTIs by \sw{cztdatasel}. Nominally, \sw{cztgtigen} discards data when the primary target is occulted by Earth. However, as the detector is sensitive to the entire sky, we include this data in our analyses. We then chose initial thresholds (see Table~\ref{tab:param})  for flagging noisy modules and pixels in \sw{cztpixclean} to ensure that transients are not suppressed.

\begin{deluxetable}{ll}
\centering
\tablecaption{Parameters used for CZTI search\label{tab:param}}
\tablehead{\colhead{Parameter} & \colhead{Value}}
\startdata
CZT module clean threshold & 1000\tablenotemark{a}\\
CZT pixel clean threshold & 100\tablenotemark{b}\\
Energy range & 20 -- 200~keV\\
Energy bin (qualitative search) & 20~keV \\
Background de-trending & {20~s Savitzky Golay filter}\\
Search window (\tsearch) & $-10$~s to $+10$~s \\
Search timescales (\tbin) & 0.01~s, 0.1s, 1s 
\enddata
\tablenotetext{a}{Modules with more than 1000 counts in one second are flagged as noisy, and suppressed for that one second duration. The typical count rates in each module are $\sim 5-10$ counts per second.}
\tablenotetext{b}{Pixels with more than 100 counts in one second are flagged as noisy, and suppressed for that one second duration.}
\end{deluxetable}

\subsection{Qualitative searches}\label{subsec:qualitative}
After the data are prepared in this manner, we first undertake a qualitative search for FRB counterparts. We select a \tsearch=20~s window centered on the de-dispersed time of the FRB. We note that the uncertainties in the reported times of the FRBs are much smaller than our search window. For instance, uncertainty in time corresponding to DM error of 0.1~$\mathrm{pc\,cm^{-3}}$ at a frequency of 800~MHz (bottom of the band for UTMOST) will be 0.65~ms, while that of due to \asat's~positional error of 1 deg will be 0.78~ms.

\begin{figure}[htb]
\centering
\includegraphics[scale=0.45]{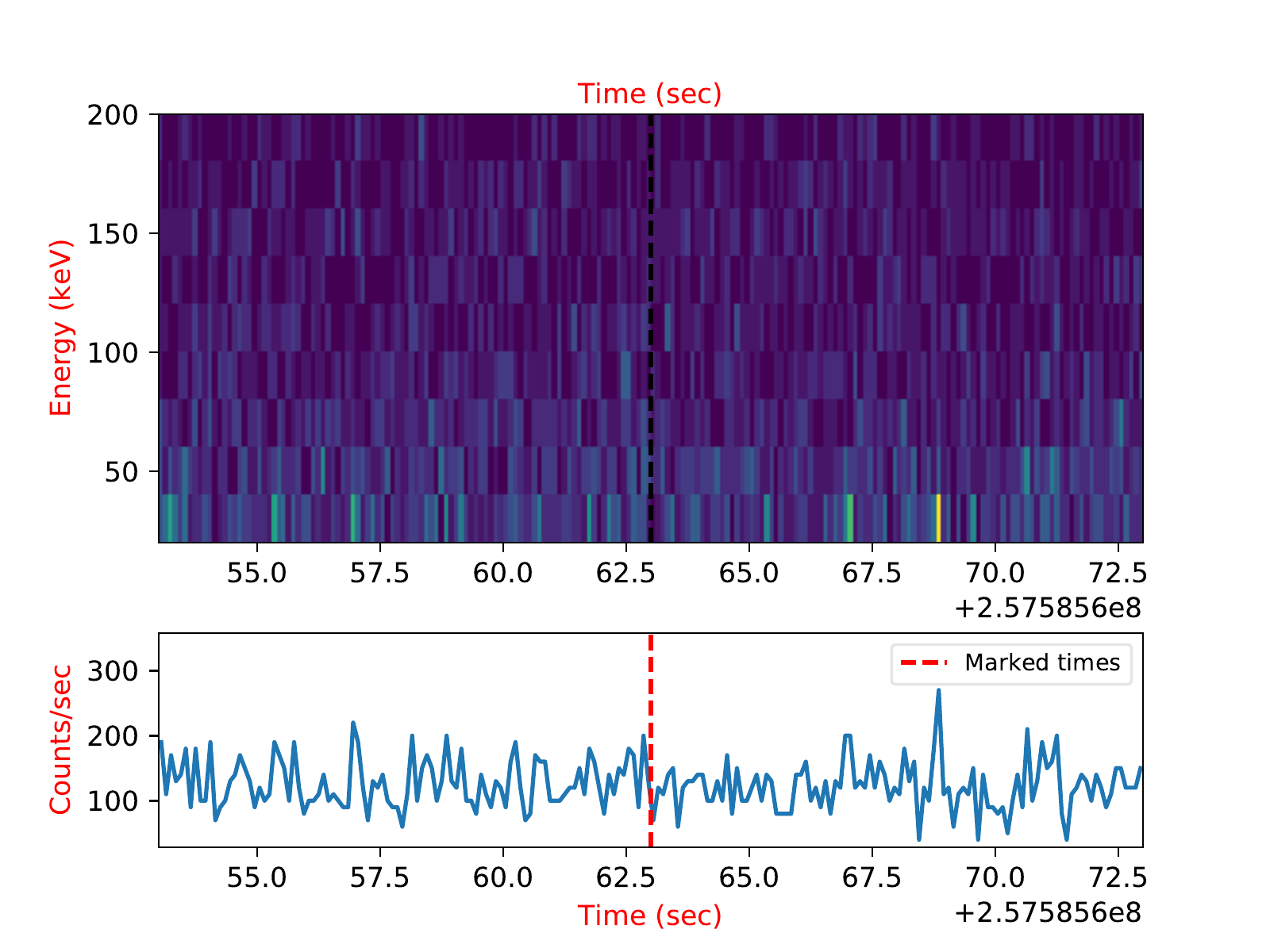}
\includegraphics[scale=0.45]{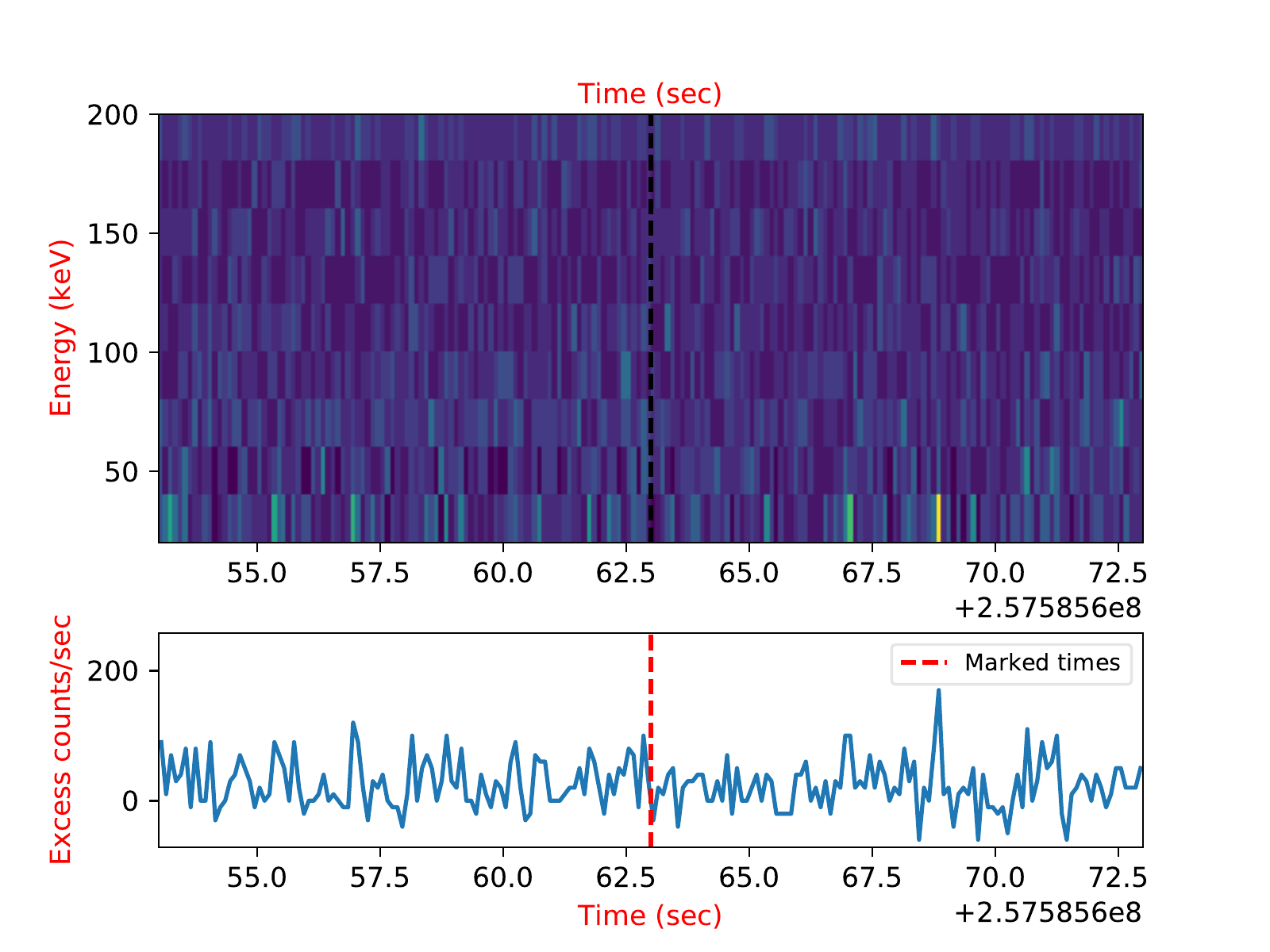}
\includegraphics[scale=0.45]{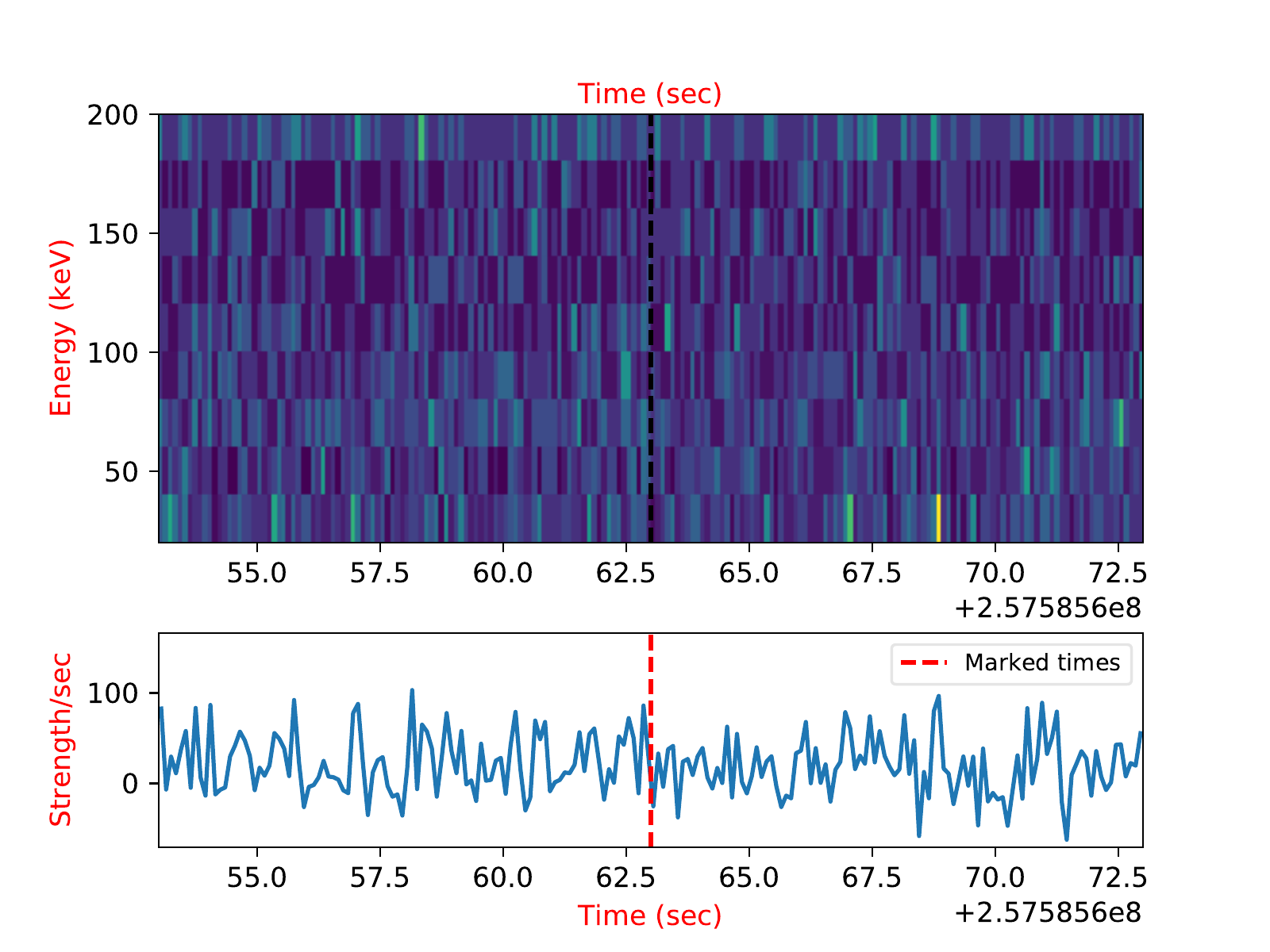}
\caption{Spectrograms corresponding temporally to search windows centered around arrival time at infinite frequency post dispersive delay correction (TOA) of FRB 180301, marked by a dashed red line. \textit{Top Panel:} Spectrogram for CZTI quadrant~C. \textit{Middle Panel:} Median subtracted spectrogram. \textit{Bottom Panel:} Median-subtracted and normalized spectrogram. See \S\ref{subsec:qualitative} for more details. All times are UTC, without a barycentric correction.
}
\label{fig:Light_Curves}
\end{figure}

Each of the four independent, identical quadrants of CZTI are treated as a separate instrument. We create two dimensional spectrograms (time-energy plots) for the data of each quadrant by binning the data in 20~keV bins in the nominal CZTI energy range of 20--200~keV, and in temporal bins of width \tbin\ (Figure~\ref{fig:Light_Curves}, top). The spectrograms are plotted and visually inspected for excess emission in the \tsearch\ window. In this work, we conducted the searches at three timescales: \tbin = 10~ms, 100~ms, and 1~s. We do not search at 1~ms timescale: given the effective area of CZTI, the count rate required to ensure at most a single false positive in the \tsearch\ window will be an order of magnitude higher than the count rate observed in the succeeding \tbin\ (10~ms) which means the source has to extremely bright to be able to get detectable at 1~ms timescale. These spectrograms are dominated by the typical ``background'' spectrum of CZTI, which in our case also includes photons from the on-axis source at the instant of the search. Hence, we also create median-subtracted spectrograms by calculating a median count rate for each energy bin and subtracting it from the instantaneous counts at that energy in each time bin (Figure~\ref{fig:Light_Curves}, middle). As a last step, we enhance outliers in each energy bin by dividing the light curve at that energy with its standard deviation calculated over the entire \tsearch\ window (Figure~\ref{fig:Light_Curves}, bottom). We visually inspect these spectrograms for any signs of enhanced X-ray emission in the \tsearch\ window. For all publicly available FRBs where coincident CZTI data were available (Table \ref{tab:frb_obs_limits}), we did not detect any X-ray candidates.

%The disperion delay delay between a X-ray and a radio emission is given by $t_\mathrm{DM}\footnote{$t_\mathrm{DM}$ is in msec,DM has the units of $pc\,cm^{-3}$ and $\nu$ is in GHz} =  4.1488\, \times (\mathrm{DM}) \times (\nu)^{-2}$ msec. A dedicated search, summarized in the subsequent subsection, is then carried to find any X-ray counterpart detection corresponding temporally with an FRB detection.

\subsection{Flux limits}\label{subsec:fluxlim}

Calculation of upper limits for our X-ray non-detections involves three steps. First, we quantify the cut-off count rates above which an FRB would have been detected at high confidence. Second, we calculate the effective area of CZTI in the direction of each FRB. Finally, we assume a spectral model for the X-ray emission and convert the count rate limits into flux upper limits.

The figure of merit we choose for upper limits is the false alarm rate: thus allowing us to put upper limits with a certain confidence. For a candidate to be considered a ``detection'', we require coincident detection in all four CZTI quadrants\footnote{We note that there are physically plausible scenarios in which a signal may be detected in only two or three quadrants, but these scenarios require more advanced treatment of data beyond the scope of the current work.}. We select the minimum counts requirement as the point where the probability of accidentally getting those many counts in a quadrant is $\qfar = 10\%$. Since the four quadrants are independent, the combined false alarm rate is $\qfar^4 =10^{-4}$. Hence, we can state that the counts (and flux) from any FRB are lower than our calculated cutoff rates with 99.99\% confidence.

If event rates in CZTI were Poisson, we could directly calculate the statistical significance and the false alarm probability of each outlier in the light curve. However, observations have shown that is not true: the data  deviate from a Poisson distribution. Hence, we estimate the false alarm rate by actually measuring the behaviour of the count rate in neighbouring data. We take all data from five orbits before and after the instant of the FRB\footnote{The FRB search window is excluded from this background estimation.}, typically amounting to about 40~ks of data. We process these comparison orbits in the same way as the FRB orbit (\S~\ref{subsec:reduction}). Instead of visual inspection, we now take a more quantitative route. We briefly describe data detrending and cut-off rate estimation here; a more detailed explanation alongside plots is provided in Appendix \ref{sec:extra_notes}. 

We calculate light curves for all comparison in the complete energy band (20 -- 200~keV), binned at the appropriate \tbin. The background count rate is variable on timescales of several minutes based on the position of the satellite around Earth. This slow variation in the background is subtracted off by using a second order Savitzky-Golay filter with a width of 20 seconds. We calculate histograms of count rates from these de-trended light curves, and measure a cutoff rate such that the probability of randomly getting counts above the cutoff rate in a \tsearch\ window is \qfar. We then create light curves for the \tsearch\ window centered on the FRB, and check if the de-trended counts exceed the cutoff rate for that quadrant. As discussed above, a detection would consist of the count rates exceeding the cutoff rates simultaneously in all quadrants. In this study, we repeated this process at all three binning timescales. Based on this criterion, we have X-ray non-detections for all 41 FRBs in our sample.

The sensitivity of the satellite to a burst depends on the location of the burst in the satellite frame of reference. Various satellite elements absorb and scatter incident photons, reducing the number of photons reaching the detector. The sensitivity is highest for on-axis sources, and lowest for bursts that are ``seen'' through the satellite body. The CZTI team has developed a GEANT4-based mass model of the satellite\footnote{Details of the mass model will be reported elsewhere --- Mate et al., in prep.} to calculate the energy- and direction-dependent effective area and response of the satellite. For all FRBs in our sample, we take the coordinates from FRBCAT and convert them into the satellite frame based on the orientation of the satellite at that instant. We then run mass model simulations to calculate the satellite response for each FRB. We note that the effective area and photon energy redistribution (ARF and RMF in X-ray parlance) obtained from the mass model do not change considerably over the uncertainties in the positions of FRBs - hence our inferred upper limits remain reliable.

The last step is the conversion of our cutoff count rates to flux and fluence upper limits for X-ray emission from the FRBs. We start by assuming a simple power law spectrum, $N(E) \propto E^{\Gamma}$, with photon power law index $\Gamma = -1$. The corresponding upper limits on hard X-ray flux are reported in \S\ref{sec:results}. We then estimate $\Gamma_\mathrm{max}$, the maximal value of the power law index that is consistent with our assumptions. To do this, we assume a single power law spectrum from radio to hard X-ray energies, and estimate its power law index from our flux limits. This new power law index is used to calculate a new flux limit using our count rate cutoff and the mass model response files. The process is repeated until $\Gamma_\mathrm{max}$ converges. Lastly, we use this $\Gamma_\mathrm{max}$ value to also calculate limits on the gamma-ray to radio flux ratio that can be compared with past results (\S\ref{sec:results}).

\section{Results}\label{sec:results}
For this work, we limit ourselves to the time period from 2015 October 6 (the day CZTI became operational) to 2018 August 31. FRBCAT~\citep{frbcat} lists 64 FRBs in this period --- of which 16 were occulted by Earth in the CZTI frame, two were very close to the Earth limb and are ignored, while five occurred while CZTI was non-functional due to a passage of \asat\ through the South Atlantic Anomaly. Table~\ref{tab:frb_obs_limits} lists the properties of the remaining 41 FRBs that were visible to CZTI. The radio fluences for the Parkes bursts are defined assuming the burst was in the center of the beam, and hence are lower limits. We searched the CZTI data for prompt emission in the 20--200~keV band from the remaining 41 FRBs at three timescales, and did not obtain any detection. Corresponding upper limits on X-ray fluence (erg~cm$^{-2}$) assuming a simple power-law spectrum with photon index $\Gamma = -1$ are reported in Table~\ref{tab:frb_xray_limits}. The table also reports upper limits on the X-ray to radio fluence ratios ($\eta$) by self-consistently choosing a power-law slope as discussed in \S\ref{subsec:fluxlim}.

\section{Discussion}

We have set limits on the X-ray to radio fluence ratios that vary between $\sim 10^{7}$ and $\sim 10^{10}$ depending on the intrinsic brightness of the FRB, its location in the radio telescope beam and the location relative to the CZTI field of view. This approaches the range of X-ray to radio fluence ratios expected from theory ($\sim10^{4}$, \citealt{2002ApJ...580L..65L} to $10^{6}$, \citealt{2014MNRAS.442L...9L}) and also the range of observed values for pulsars. 

\citet{delaunay2016discovery} searched the \textit{Swift} BAT data for any possible $\gamma$-ray emission in the energy range 15-150 KeV and obtained fluence limit of $\mathrm{F_{\gamma}} \lesssim 10^{-6}$ $\mathrm{erg\,cm^{-2}} $ in an interval of 300~sec for a total of 4 FRBs. This corresponds to $\eta$ ($F_\mathrm{\gamma}/F_\mathrm{1.4\,GHz}$) of $\sim 10^{11}$. We note that the claimed $\gamma$-ray transient detection corresponding to FRB131104 is extremely marginal (illuminating only 2.9$\%$ of the \textit{Swift}-BAT detector). Further, the search was conducted at much  longer timescales ($T_{90}$ = 100 sec) as compared to our study, hence the claimed detection is not at odds with our limits.
%\citet{tkp16} set limits on the fluence ratio in $\gamma$-ray to radio bands for SGR 1806--20 giant flare which is detected in $\gamma$-rays but not in radio regime revealing that the upper limit $F_\mathrm{\gamma}/F_\mathrm{1.4\,GHz} \lesssim 10^{-7} - 10^{-9}\,\mathrm{erg\,cm^{-2}\,Jy^{-1}\,ms^{-1}}$ which implies that the corresponding lower limit $F_\mathrm{\gamma}/F_\mathrm{1.4\,GHz} \lesssim 10^{8}-10^{10}$ tends to the value of our proposed $\eta$.
%}

However, despite the strong upper limits from non-detections, it is challenging to constrain theories directly based on individual FRB observations because the intrinsic X-ray to radio fluence ratio may be significantly different from the observed ratio due to beaming effects. For instance, some models of binary neutron star mergers suggest that the X-ray/$\gamma$-ray emission would be strongly beamed (as in the case of relativistic jets from GRBs) while the radio emission is relatively isotropic \citep[see for instance,][and references therein.]{Totani:2013lia} In such cases, the lack of observed high-energy emission in an individual case can be dismissed. If the jets are highly relativistic, the emission will be strongly beamed and visible to $<1\%$ of all observers --- as happens for gamma ray bursts \citep{berger2014}. We need statistical limits on the X-ray to radio fluence ratios on hundreds of FRBs to help us constrain their emission models.

Conversely, if we expect that the radio emission is beamed while the X-ray emission is nearly isotropic (as in the case of a magnetar giant flare), it will be significantly more challenging to verify emission mechanisms. However, in terms of energetics, high-energy emission powered by compact objects with magnetar-like magnetic field strengths cannot be detected at gigaparsec distances unless they are relativistically beamed \citep{murase2017}. 

\asat\ CZTI is one of the most sensitive instruments for detecting short duration high energy transients \citep[see for example][]{2017ApJ...845..152B}.
%{\em AstroSat} CZTI can effectively search for X-ray counterparts of FRBs with its all-sky sensitivity. 
%When CHIME \citep{chimefrb2018} and other FRB surveys come online, this framework will be used to automatically search for counterparts as the FRB detection triggers are made available. 
Our limits on 41 bursts out of 64 that occurred in our search period are consistent with the operational expectations of being sensitive to about half the events, the rest being lost to Earth-occultation and SAA transits. \asat\ continuously records time-tagged photon data which can be used to search for FRB counterparts in ground processing. \asat\ also has the advantage of  being sensitive to the entire sky not obstructed by earth: similar to {\em Fermi}-LAT but several times larger than the field of view of {\em Swift}-BAT. This wide-field hard X-ray sensitivity of CZTI will be very useful in the future as the rate of FRB detections increases with facilities such as CHIME \citep{chimefrb2018}, ASKAP \citep{2009ASPC..407..446J}, HIRAX \citep{2016SPIE.9906E..5XN} and SKA \citep{2015arXiv150104076M}.
\acknowledgments
We thank E. Aarthy for helpful discussions in flux calculations. This publication also uses the data from the \asat\ mission of the Indian Space Research Organisation (ISRO), archived at the Indian Space Science Data Centre (ISSDC). CZT--Imager is built by a consortium of Institutes across India. The Tata Institute of Fundamental Research, Mumbai, led the effort with instrument design and development. Vikram Sarabhai Space Centre, Thiruvananthapuram provided the electronic design, assembly and testing. ISRO Satellite Centre (ISAC), Bengaluru provided the mechanical design, quality consultation and project management. The Inter University Centre for Astronomy and Astrophysics (IUCAA), Pune did the Coded Mask design, instrument calibration, and Payload Operation Centre. Space Application Centre (SAC) at Ahmedabad provided the analysis software. 
%Physical Research Laboratory (PRL) Ahmedabad, provided the polarisation detection algorithm and ground calibration. 
A vast number of industries participated in the fabrication and the University sector pitched in by participating in the test and evaluation of the payload.
The Indian Space Research Organisation funded, managed and facilitated the project.
This work utilised various software including Python, IDL, FTOOLS, C, and C++.

\vspace{5mm}
\facilities{AstroSat(CZTI)} 

\software{AstroPy \citep{2013A&A...558A..33A}, Python, IDL, FTOOLS \citep{1995ASPC...77..367B}, C,  C++}
\vspace{0.5cm}
\appendix

\section{Estimating Cut off count rate}\label{sec:extra_notes}
\begin{figure}[b]
\centering
\includegraphics[scale=0.5]{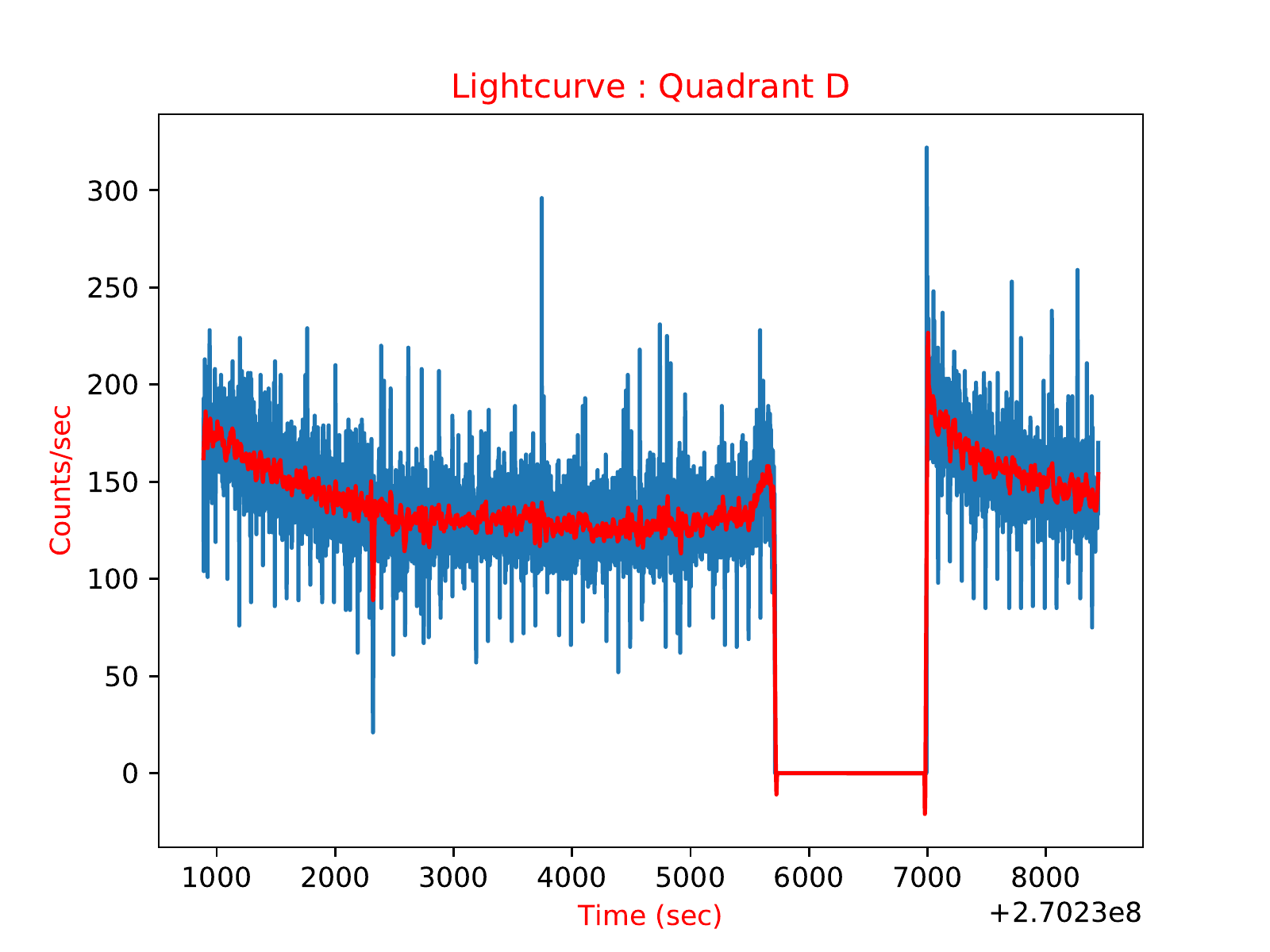}
\includegraphics[scale=0.5]{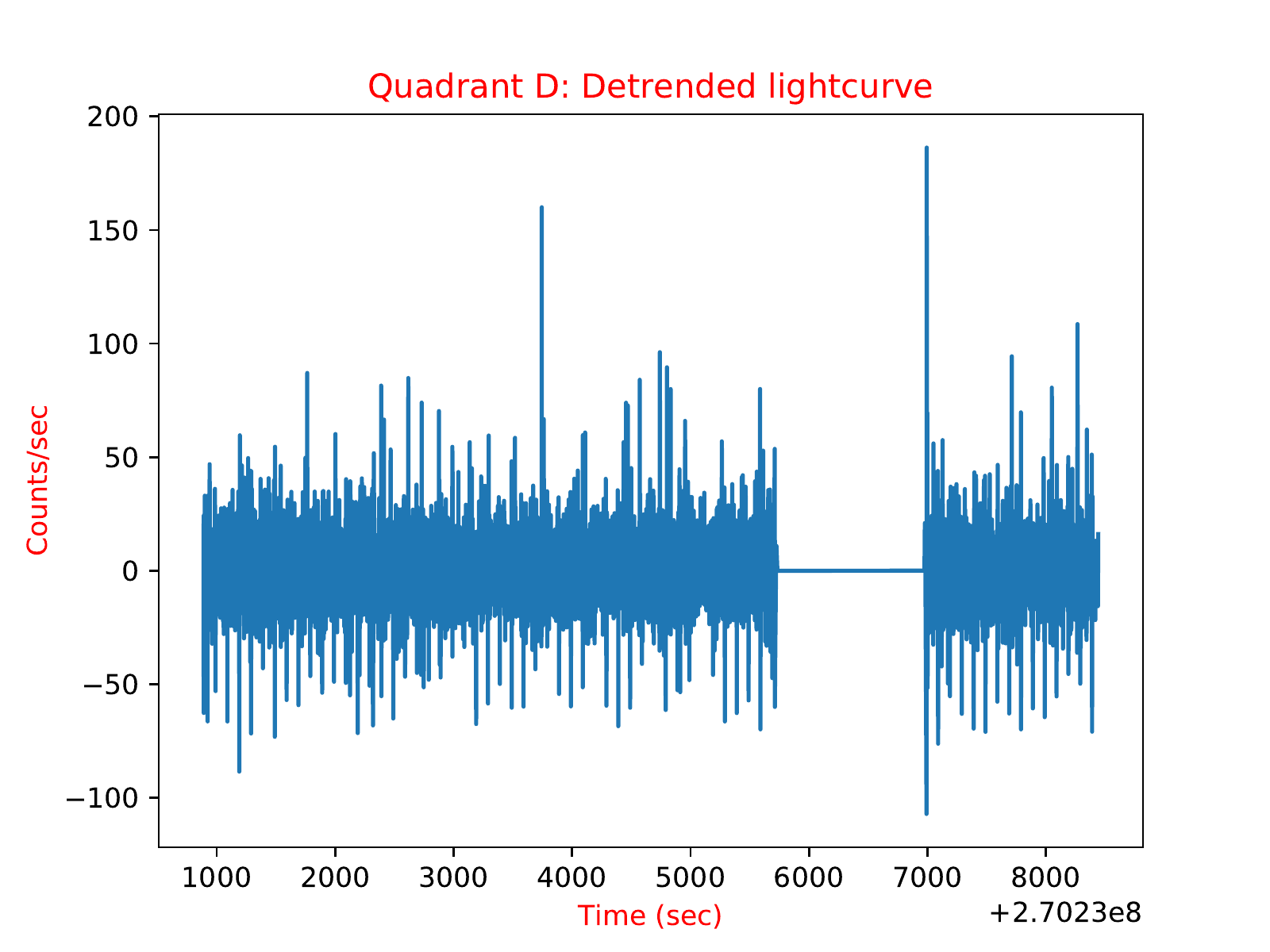}\\
\includegraphics[scale=0.5]{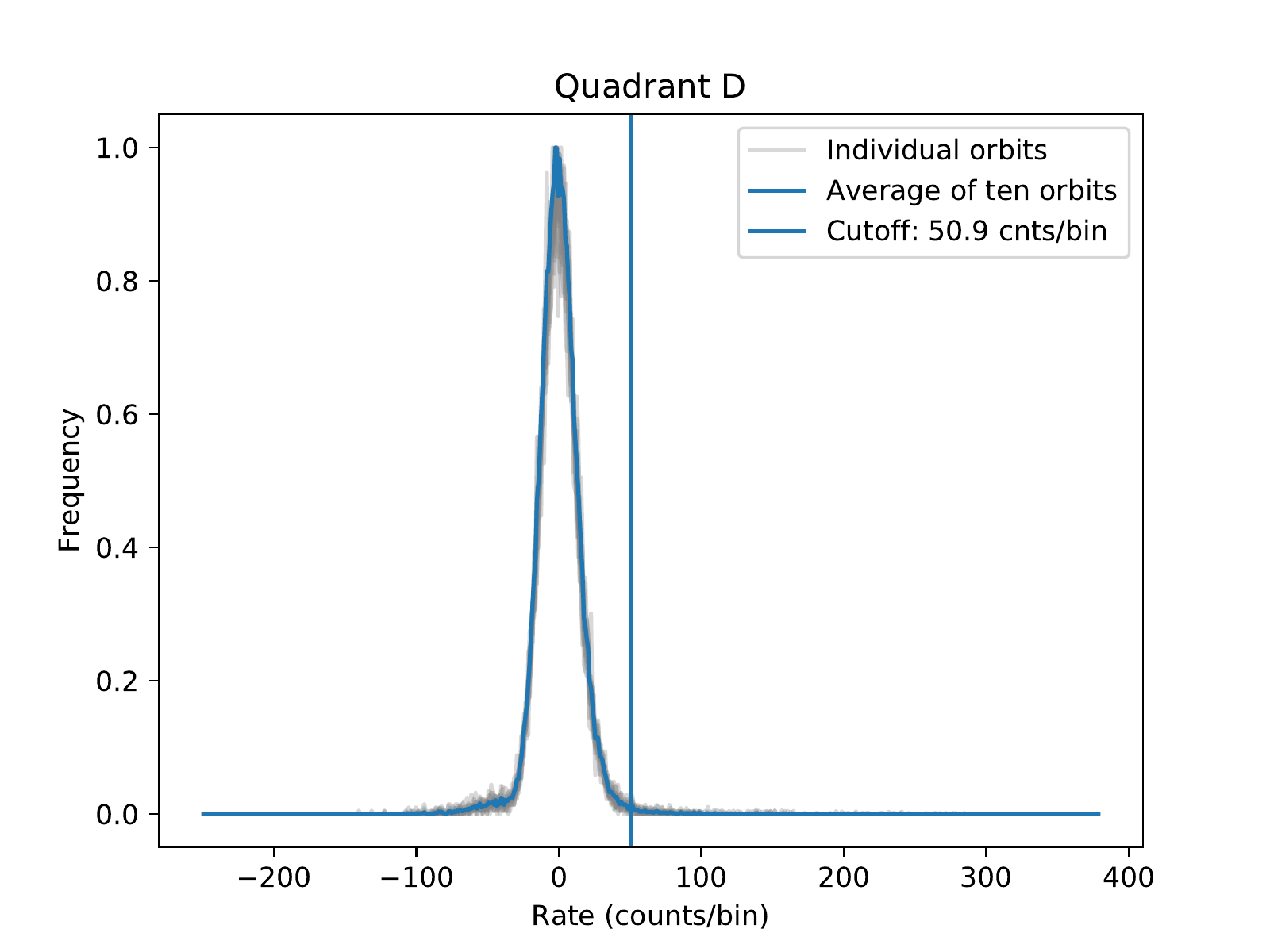}\\
\includegraphics[scale=0.5]{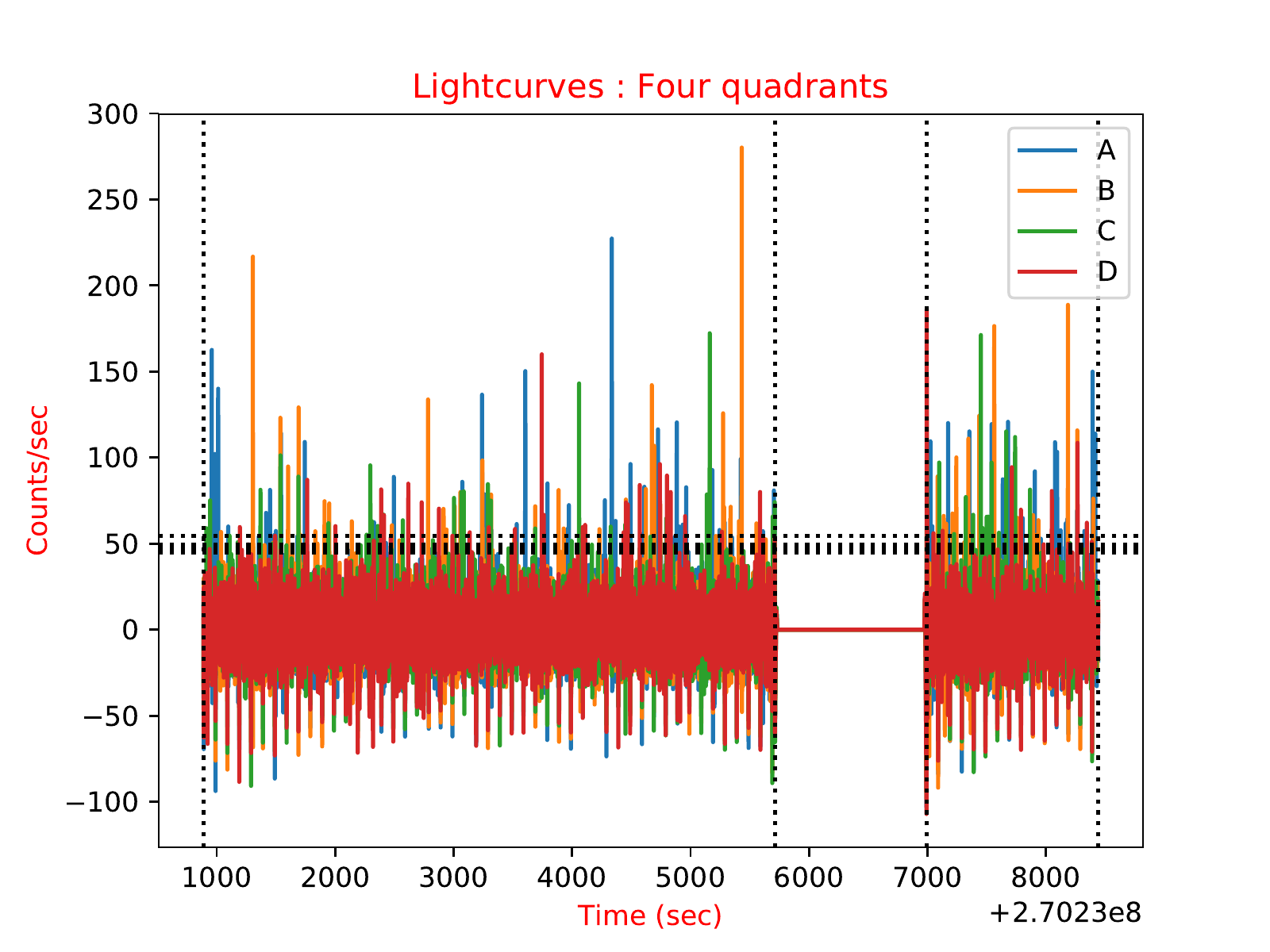}
\includegraphics[scale=0.5]{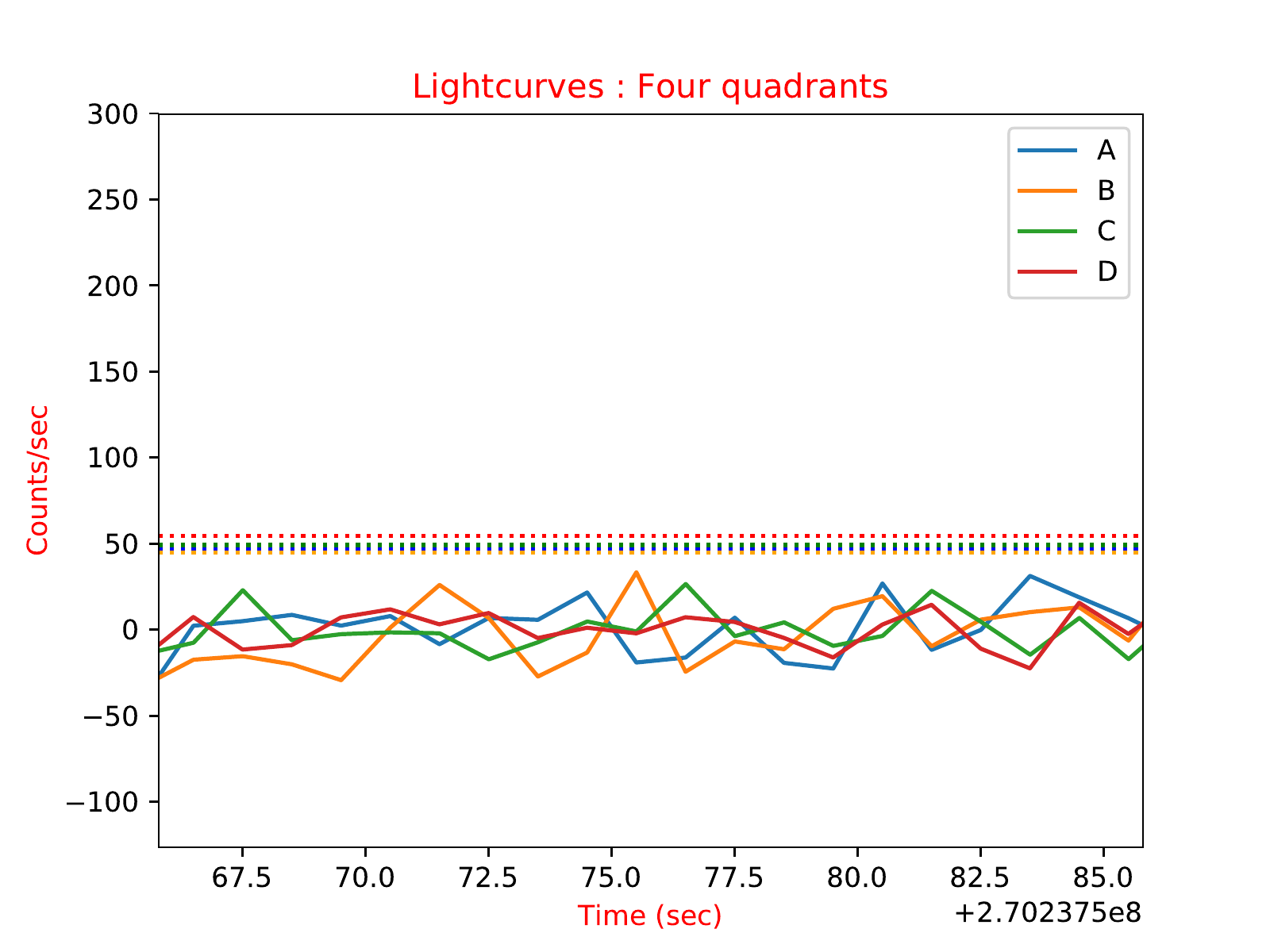}
\caption{\emph{Top Panel:} Raw light curve at 1-s binning for one of the four detector quadrants of CZTI. The slowly varying trend due to \asat's orbit and the data gap during SAA passage is visible. \emph{Middle Panel:}Normalized histogram of the light curve for all the ten individual neighbouring orbits (grey curve) and the resultant average histogram (blue) . \emph{Bottom Panel: Left} De-trended light curve for the data from the top panel. The vertical dashed lines show data segments that are used. The horizontal dashed lines shows the estimated cut-off count-rate of . \emph{Bottom Panel: Right} Light curve with quadrant-wise cut-off count rates in the search interval.
} 
\label{fig:bkg_sub}
\end{figure}

Here we describe our method for detrending the data and calculating the cut-off rate. To calculate the cut-off count rate, we choose 5 consecutive orbits each, both before and after the event orbit, excluding the orbit of interest. Figure \ref{fig:bkg_sub} (\emph{Top Panel, left}) shows the light curve of an entire orbit of one of the quadrants (Quad D) binned at 1 sec.  The slow variation in the counts reflects the motion of satellite in its orbit while the data gap is due to satellite's passage through SAA. We use a second order Savitzky-Golay filter to estimate this background variation and detrend the light curve. The red solid curve in figure \ref{fig:bkg_sub} \emph{(Top Panel, left)} shows the estimation of this slow variation using a Savitzky-Golay filter. Figure \ref{fig:bkg_sub} (\emph{Top Panel, right}) depicts the 1-s binned light curve after subtracting this slow variation. The average histogram of the count-rates of all 10 neighbouring orbits is used to estimate the cut-off rate, which is quantified by the parameter \emph{confidence}. 
Given the time span of the search interval \tsearch, the binning time \tbin\ and the False Alarm Rate FAR, the chance of detecting a false positive is 1 in (\tsearch/\tbin/FAR); hence \emph{confidence} is 1 - (\tsearch/\tbin/FAR). The cut-off rate is chosen based on this required confidence, from the average histogram and is independently estimated for each binning time. It can be noted that the false alarm rate chosen for this analysis is 0.1 per quadrant for a time span of 20s (since the transient is short-lived $\sim$ms) . The four quadrants of CZTI are independent and so the probability of getting a temporally coincident false positive in all the quadrants is $10^{-4}$ in the search interval (20s). The orbital period of \asat\ is $\sim$ 6000s; hence there can be $\sim 30$ false positives per quadrant in an orbit (Figure \ref{fig:bkg_sub} \emph{Bottom Panel, left}). However, the requirement of temporal coincidence across quadrants removes such false positives. Figure \ref{fig:bkg_sub} (\emph{Bottom Panel, right}) shows the light curve in a 20s window around the arrival time of FRB, with cut off rates for individual quadrants marked. We see no evidence for any temporally coincident prompt emission in all the quadrants above the background level.

\newpage

\begin{deluxetable*}{ccccccccccc}
\def\arraystretch{1.25}
  \centering
  \tablecolumns{9} 
  \tablecaption{Observed Parameters of Radio Bursts. \label{tab:frb_obs_limits}}
  \tablewidth{\textwidth}
  \tabletypesize{\footnotesize}
  \tablehead{
\colhead{Name} &
\colhead{Time} &
\multicolumn{2}{c}{Coordinates (J2000)} &
\colhead{Radio Telescope} &
\colhead{Central frequency} &
\colhead{Bandwidth} &
\colhead{$S_\mathrm{radio}\tablenotemark{a}$} &
\colhead{FWHM} &
\colhead{$F_\mathrm{radio}\tablenotemark{a}$} \\
\colhead{}&
\colhead{UTC}&
\colhead{RA}&
\colhead{Dec}&
\colhead{}&
\colhead{MHz}&
\colhead{MHz}&
\colhead{Jy}&
\colhead{ms}&
\colhead{Jy-ms} &
  }
  \startdata
FRB190806 & 17:07:58.0 & 00:02:21.38 & $-$07:34:54.6  & UTMOST & 835.0 & 31.25 & 3.91 & 11.96 & 46.8 \\  
FRB190714 & 05:37:12.901 & 12:15.9 & $-$13:00  & ASKAP & 1297.0 & 336.0 & 4.7 & 1.7 & 8.0 \\
FRB190711 & 01:53:41.100 & 21:56 & $-$80:23  & ASKAP & 1297.0 & 336.0 & 4.11 & 9.0 & 28.0 \\
FRB190523 & 06:05:55.815 & 13:48:15.6 & +72:28:11  & DSA$-$10 & 1405.0 & 125.0 & 666.67 & 0.42 & 280.0 \\  
FRB190322 & 07:00:12.3 & 04:46:14.45 & $-$66:55:27.8  & UTMOST & 835.0 & 31.25 & 11.85 & 1.35 & 16.0 \\  
FRB181228 & 13:48:50.100 & 06:09:23.64 & $-$45:58:02.4 & UTMOST & 835.0 & 31.25 & 19.23 & 1.24 & 23.85 \\  
FRB181017 & 10:24:37.400 & 22:05:54.82 & $-$08:50:34.22 & UTMOST & 835.0 & 31.25 & 161 & 0.32 & 51.52 \\
FRB180817.J1533+42 & 01:49:20.202 & 15:33 & +42:12 & CHIME/FRB & 600.0 & 400.0 & 70.27 & 0.37 & 26.0 \\
FRB180814.J1554+74 & 14:20:14.440 & 15:54 & +74:01 & CHIME/FRB & 600.0 & 400.0 & 138.89 & 0.18 & 25.0 \\
FRB180812.J0112+80 & 11:45:32.872 & 01:12 & +80:47 & CHIME/FRB & 600.0 & 400.0 & 14.4 & 1.25 & 18.0 \\
FRB180810.J1159+83 & 22:40:42.493 & 11:59 & +83:07 & CHIME/FRB & 600.0 & 400.0 & 60.71 & 0.28 & 17.0 \\
FRB180806.J1515+75 & 14:13:03.107 & 15:15 & +75:38 & CHIME/FRB & 600.0 & 400.0 & 34.78 & 0.69 & 24.0 \\
FRB180801.J2130+72 & 08:47:14.793 & 21:30 & +72:43 & CHIME/FRB & 600.0 & 400.0 & 54.9 & 0.51 & 28.0 \\
FRB180730.J0353+87 & 03:37:25.937 & 03:53 & +87:12 & CHIME/FRB & 600.0 & 400.0 & 119.05 & 0.42 & 50.0 \\
FRB180729.J0558+56 & 17:28:18.258 & 05:58 & +56:30 & CHIME/FRB & 600.0 & 400.0 & 112.5 & 0.08 & 9.0 \\
FRB180727.J1311+26 & 00:52:04.474 & 13:11 & +26:26 & CHIME/FRB & 600.0 & 400.0 & 17.95 & 0.78 & 14.0 \\
FRB180725.J0613+67 & 17:59:32.813 & 06:13 & +67:04 & CHIME/FRB & 600.0 & 400.0 & 38.71 & 0.31 & 12.0 \\
FRB 180528 & 04:24:00.9 & 06:38:48.7 & $-$49:53:59 & UTMOST & 835.0 & 32 & 15.75 & 2.0 & 32 \\
FRB180525 & 15:19:06.515 & 14:40 & $-$02:12 & ASKAP & 1297.0 & 336.0 & 78.9 & 3.8 & 299.82 \\
FRB180430 & 10:00:35.700 & 06:51 & $-$09:57 & ASKAP & 1297.0 & 336.0 & 147.5 & 1.2 & 177.0 \\
FRB180324 & 09:31:46.706 & 06:16 & $-$34:47 & ASKAP & 1297.0 & 336.0 & 16.5 & 4.3 & 70.95 \\
FRB180315 & 05:05:30.985 & 19:35 & $-$26:50 & ASKAP & 1297.0 & 336.0 & 23.3 & 2.4 & 55.92 \\
FRB 180311 & 04:11:54.800 & 21:31:33.42 & $-$57:44:26.7 & Parkes & 1352.0 & 338.381 & 0.2 & 12.0 & 2.4 \\
FRB 180301 & 07:34:19.760 & 06:12:43.4 & 04:33:44.8 & Parkes & 1352.0 & 338.381 & 0.5 & 3.0 & 1.5 \\
FRB180212 & 23:45:04.399 & 14:21 & $-$03:35 & ASKAP & 1297.0 & 336.0 & 53.0 & 1.81 & 95.93 \\
FRB180130 & 04:55:29.993 & 21:52.2 & $-$38:34 & ASKAP & 1297.0 & 336.0 & 23.1 & 4.1 & 94.71 \\
FRB180119 & 12:24:40.747 & 03:29.3 & $-$12:44 & ASKAP & 1297.0 & 336.0 & 40.7 & 2.7 & 109.89 \\
FRB180110 & 07:34:34.959 & 21:53.0 & $-$35:27 & ASKAP & 1297.0 & 336.0 & 128.1 & 3.2 & 409.92 \\
FRB171213 & 14:22:40.467 & 03:39 & $-$10:56 & ASKAP & 1297.0 & 336.0 & 88.6 & 1.5 & 132.9 \\
FRB 171209 & 20:34:23.500 & 15:50:25 & $-$46:10:20 & Parkes & 1352.0 & 338.381 & 0.92 & 2.5 & 2.3 \\
FRB171020 & 10:27:58.598 & 22:15 & $-$19:40 & ASKAP & 1297.0 & 336.0 & 117.6 & 3.2 & 376.32 \\
FRB171019 & 13:26:40.097 & 22:17.5 & $-$08:40 & ASKAP & 1297.0 & 336.0 & 40.5 & 5.4 & 218.7 \\
FRB171003 & 04:07:23.781 & 12:29.5 & $-$14:07 & ASKAP & 1297.0 & 336.0 & 40.5 & 2.0 & 81.0 \\
FRB170906 & 13:06:56.488 & 21:59.8 & $-$19:57 & ASKAP & 1297.0 & 336.0 & 29.6 & 2.5 & 74.0 \\
FRB 170827 & 16:20:18.000 & 00:49:18.66 & $-$65:33:02.3 & UTMOST & 835.0 & 32 & 50.3 & 0.4 & 19.87 \\
FRB170707 & 06:17:34.354 & 02:59 & $-$57:16 & ASKAP & 1297.0 & 336.0 & 14.8 & 3.5 & 51.8 \\
FRB170606 & 10:03:27.000 & 5:34:0.0 & 41:45:0.0 & Pushchino & 111.0 & 2.5 & 0.54 & 3300.0 & 1782.0 \\
FRB170428 & 18:02:34.700 & 21:47 & $-$41:51 & ASKAP & 1320.0 & 336.0 & 7.7 & 4.4 & 33.88 \\
FRB170416 & 23:11:12.799 & 22:13 & $-$55:02 & ASKAP & 1320.0 & 336.0 & 19.4 & 5.0 & 97.0 \\
%FRB170922 & 11:22:23.400 & 21:29:50.61 & $-$07:59:40.49 & UTMOST & 835.0 & 32 & 2.3 & 26.0 & 59.8 \\
FRB 160608 & 03:53:01.088 & 07:36:42 & $-$40:47:52 & UTMOST & 843.0 & 16\tablenotemark{b}  & 4.3 & 9.0 & 38.7 \\
FRB 151230 & 16:15:46.525 & 09:40:50 & $-$03:27:05 & Parkes & 1352.0 & 338.381 & 0.42 & 4.4 & 1.9 \\
    \enddata
%\tablenotetext{a}{$f_\mathrm{cen}$ is the central frequency}
\tablenotetext{a}{$S_\mathrm{radio}$ is radio flux and $F_\mathrm{radio}$ is radio fluence of the burst.}
%\tablenotetext{c}{$F_\mathrm{radio}$ is radio fluence}
\tablenotetext{b}{The value reported is assumed to be 16 on the basis of previous detection (FRB 160317) since the actual value is missing from FRBCAT.}
\end{deluxetable*}

\newpage

\startlongtable
\begin{deluxetable*}{lccccccccccc}
\def\arraystretch{1.45}
  \centering
  \tablecolumns{9} 
  \tablecaption{CZTI fluence limits on X-rays from FRBs. \label{tab:frb_xray_limits}}
  \tablewidth{0pt}
  \tabletypesize{\footnotesize}
  \tablehead{
\colhead{Name} &
\colhead{Radio Flux Density} &
\colhead{Radio Fluence} &
\colhead{tbin} &
\colhead{$\Gamma_{max}$} &
\multicolumn{2}{c}{X-ray fluence} &
\multicolumn{2}{c}{$\eta / 10^{9} = \frac{F_{X-ray}}{F_{Radio}} / 10^{9}$} \\
\colhead{(Reference to original detection)}&
\colhead{Jy}&
\colhead{Jy-ms}&
\colhead{s}&
\colhead{}&
\multicolumn{2}{c}{$\mathrm{erg~cm^{-2}}$}&
\multicolumn{2}{c}{} \\
\colhead{}&
\colhead{}&
\colhead{}&
\colhead{}&
\colhead{}&
\colhead{$\Gamma = -1$}&
\colhead{$\Gamma = \Gamma_{max}$}&
\colhead{$\Gamma = -1$}&
\colhead{$\Gamma = \Gamma_{max}$}&
  }
  \startdata
FRB190806 & 3.91 & 46.8 & 0.01 & $-$1.19 & 1.6e$-$07 & 1.65e$-$07 & 0.34 & 0.35 \\
\citep{FRB190806} &  &  & 0.1 & $-$1.25 & 3.67e$-$07 & 3.84e$-$07 & 0.78 & 0.82 \\
 &  &  & 1.0 & $-$1.33 & 5.69e$-$07 & 6.03e$-$07 & 1.21 & 1.29 \\
FRB190714 & 4.7 & 8.0 & 0.01 & $-$1.24 & 7.38e$-$08 & 7.47e$-$08 & 0.92 & 0.93 \\
\citep{FRB190714} &  &  & 0.1 & $-$1.3 & 1.67e$-$07 & 1.69e$-$07 & 2.08 & 2.11 \\
 &  &  & 1.0 & $-$1.38 & 2.72e$-$07 & 2.76e$-$07 & 3.4 & 3.45 \\
FRB190711 & 4.1 & 28.0 & 0.01 & $-$1.16 & 4.33e$-$07 & 4.44e$-$07 & 1.55 & 1.59 \\
\citep{FRB190711} &  &  & 0.1 & $-$1.22 & 9.72e$-$07 & 1.01e$-$06 & 3.47 & 3.6 \\
 &  &  & 1.0 & $-$1.3 & 1.55e$-$06 & 1.64e$-$06 & 5.55 & 5.85 \\
FRB190523 & 666.7 & 280.0 & 0.01 & $-$1.42 & 1.53e$-$07 & 1.61e$-$07 & 0.05 & 0.06 \\
\citep{FRB190523} &  &  & 0.1 & $-$1.48 & 3.51e$-$07 & 3.74e$-$07 & 0.13 & 0.13 \\
 &  &  & 1.0 & $-$1.56 & 5.56e$-$07 & 6.02e$-$07 & 0.2 & 0.21 \\
FRB190322 & 11.8 & 16.0 & 0.01 & $-$1.23 & 2.15e$-$07 & 2.23e$-$07 & 1.35 & 1.39 \\
\citep{FRB190322} &  &  & 0.1 & $-$1.29 & 4.83e$-$07 & 5.07e$-$07 & 3.02 & 3.17 \\
 &  &  & 1.0 & $-$1.37 & 7.53e$-$07 & 8.02e$-$07 & 4.7 & 5.01 \\
FRB181228 & 19.2 & 23.8 & 0.01 & $-$1.28 & 9.74e$-$08 & 9.6e$-$08 & 0.41 & 0.4 \\
\citet{ASKAP5} &  &  & 0.1 & $-$1.36 & 1.38e$-$07 & 1.36e$-$07 & 0.58 & 0.57 \\
 &  &  & 1.0 & $-$1.44 & 2.08e$-$07 & 2.04e$-$07 & 0.87 & 0.85 \\
FRB181017 & 161.0 & 51.5 & 0.01 & $-$1.39 & 6.8e$-$08 & 6.51e$-$08 & 0.13 & 0.13 \\
\citet{ASKAP5} &  &  & 0.1 & $-$1.47 & 9.7e$-$08 & 9.2e$-$08 & 0.19 & 0.18 \\
 &  &  & 1.0 & $-$1.55 & 1.45e$-$07 & 1.36e$-$07 & 0.28 & 0.26 \\
FRB180817.J1533+42 & 70.3 & 26.0 & 0.01 & $-$1.3 & 2.17e$-$07 & 2.31e$-$07 & 0.83 & 0.89 \\
\citep{CHIME_FRBs} &  &  & 0.1 & $-$1.36 & 4.84e$-$07 & 5.22e$-$07 & 1.86 & 2.01 \\
 &  &  & 1.0 & $-$1.43 & 7.57e$-$07 & 8.33e$-$07 & 2.91 & 3.2 \\
FRB180814.J1554+74 & 138.9 & 25.0 & 0.01 & $-$1.35 & 1.18e$-$07 & 1.21e$-$07 & 0.47 & 0.48 \\
\citep{CHIME_FRBs} &  &  & 0.1 & $-$1.41 & 2.71e$-$07 & 2.78e$-$07 & 1.08 & 1.11 \\
 &  &  & 1.0 & $-$1.49 & 4.31e$-$07 & 4.46e$-$07 & 1.72 & 1.78 \\
FRB180812.J0112+80 & 14.4 & 18.0 & 0.01 & $-$1.25 & 1.26e$-$07 & 1.29e$-$07 & 0.7 & 0.72 \\
\citep{CHIME_FRBs} &  &  & 0.1 & $-$1.31 & 2.91e$-$07 & 3.01e$-$07 & 1.62 & 1.67 \\
 &  &  & 1.0 & $-$1.39 & 4.41e$-$07 & 4.62e$-$07 & 2.45 & 2.57 \\
FRB180810.J1159+83 & 60.7 & 17.0 & 0.01 & $-$1.2 & 1.85e$-$06 & 2e$-$06 & 10.89 & 11.74 \\
\citep{CHIME_FRBs} &  &  & 0.1 & $-$1.26 & 4.29e$-$06 & 4.74e$-$06 & 25.24 & 27.89 \\
 &  &  & 1.0 & $-$1.33 & 6.91e$-$06 & 7.89e$-$06 & 40.66 & 46.39 \\
FRB180806.J1515+75 & 34.8 & 24.0 & 0.01 & $-$1.27 & 2.22e$-$07 & 2.30e$-$07 & 0.93 & 0.96 \\
\citep{CHIME_FRBs} &  &  & 0.1 & $-$1.33 & 5.1e$-$07 & 5.34e$-$07 & 2.13 & 2.23 \\
 &  &  & 1.0 & $-$1.4 & 8.1e$-$07 & 8.6e$-$07 & 3.37 & 3.58 \\
FRB180801.J2130+72 & 54.9 & 28.0 & 0.01 & $-$1.29 & 2.01e$-$07 & 2.10e$-$07 & 0.72 & 0.75 \\
\citep{CHIME_FRBs} &  &  & 0.1 & $-$1.35 & 4.58e$-$07 & 4.85e$-$07 & 1.64 & 1.73 \\
 &  &  & 1.0 & $-$1.43 & 7.21e$-$07 & 7.75e$-$07 & 2.57 & 2.77 \\
FRB180730.J0353+87 & 119.0 & 50.0 & 0.01 & $-$1.31 & 2.80e$-$07 & 2.875e$-$07 & 0.56 & 0.58 \\
\citep{CHIME_FRBs} &  &  & 0.1 & $-$1.37 & 6.42e$-$07 & 6.64e$-$07 & 1.28 & 1.33 \\
 &  &  & 1.0 & $-$1.44 & 9.96e$-$07 & 1.04e$-$06 & 1.99 & 2.08 \\
FRB180729.J0558+56 & 112.5 & 9.0 & 0.01 & $-$1.3 & 3.34e$-$07 & 3.52e$-$07 & 3.71 & 3.91 \\
\citep{CHIME_FRBs} &  &  & 0.1 & $-$1.36 & 7.47e$-$07 & 7.98e$-$07 & 8.3 & 8.86 \\
 &  &  & 1.0 & $-$1.43 & 1.22e$-$06 & 1.33e$-$06 & 13.59 & 14.76 \\
FRB180727.J1311+26 & 18.0 & 14.0 & 0.01 & $-$1.21 & 4.93e$-$07 & 5.08e$-$07 & 3.52 & 3.63 \\
\citep{CHIME_FRBs} &  &  & 0.1 & $-$1.27 & 1.09e$-$06 & 1.14e$-$06 & 7.82 & 8.15 \\
 &  &  & 1.0 & $-$1.34 & 1.72e$-$06 & 1.82e$-$06 & 12.32 & 13.02 \\
FRB180725.J0613+67 & 38.7 & 12.0 & 0.01 & $-$1.25 & 3.73e$-$07 & 3.96e$-$07 & 3.11 & 3.3 \\
\citep{CHIME_FRBs} &  &  & 0.1 & $-$1.31 & 8.73e$-$07 & 9.43e$-$07 & 7.28 & 7.86 \\
 &  &  & 1.0 & $-$1.38 & 1.39e$-$06 & 1.53e$-$06 & 11.57 & 12.77 \\
FRB180528 & 15.8 & 32.0 & 0.01 & $-$1.24 & 1.21e$-$06 & 1.31e$-$06 & 3.77 & 4.1 \\
\citet{ASKAP5} &  &  & 0.1 & $-$1.3 & 2.752e$-$06 & 3.07e$-$06 & 8.6 & 9.6 \\
 &  &  & 1.0 & $-$1.38 & 4.32e$-$06 & 4.97e$-$06 & 13.49 & 15.55 \\
FRB180525 & 78.9 & 299.8 & 0.01 & $-$1.35 & 8.14e$-$08 & 8.23e$-$08 & 0.03 & 0.03 \\
\citep{ASKAP3} &  &  & 0.1 & $-$1.42 & 1.85e$-$07 & 1.87e$-$07 & 0.06 & 0.06 \\
 &  &  & 1.0 & $-$1.5 & 2.94e$-$07 & 2.99e$-$07 & 0.1 & 0.1 \\
FRB180430 & 147.5 & 177.0 & 0.01 & $-$1.32 & 2.85e$-$07 & 3.065e$-$07 & 0.16 & 0.17 \\
\citep{FRB180430} &  &  & 0.1 & $-$1.39 & 6.43e$-$07 & 7.03e$-$07 & 0.36 & 0.4 \\
 &  &  & 1.0 & $-$1.46 & 1.04e$-$06 & 1.16e$-$06 & 0.59 & 0.65 \\
FRB180324 & 16.5 & 71.0 & 0.01 & $-$1.26 & 1.59e$-$07 & 1.64e$-$07 & 0.22 & 0.23 \\
\citep{ASKAP3} &  &  & 0.1 & $-$1.32 & 3.57e$-$07 & 3.72e$-$07 & 0.5 & 0.52 \\
 &  &  & 1.0 & $-$1.4 & 5.46e$-$07 & 5.76e$-$07 & 0.77 & 0.81 \\
FRB180315 & 23.3 & 55.9 & 0.01 & $-$1.2 & 9.84e$-$07 & 1.03e$-$06 & 1.76 & 1.84 \\
\citep{ASKAP3} &  &  & 0.1 & $-$1.26 & 2.25e$-$06 & 2.39e$-$06 & 4.03 & 4.28 \\
 &  &  & 1.0 & $-$1.34 & 3.48e$-$06 & 3.77e$-$06 & 6.22 & 6.74 \\
FRB180311 & 0.2 & 2.4 & 0.01 & $-$1.02 & 5.435e$-$07 & 5.46e$-$07 & 22.64 & 22.76 \\
\citep{PARKES} &  &  & 0.1 & $-$1.08 & 1.20e$-$06 & 1.23e$-$06 & 50.17 & 51.25 \\
 &  &  & 1.0 & $-$1.16 & 2.01e$-$06 & 2.095e$-$06 & 83.68 & 87.28 \\
FRB180301 & 0.5 & 1.5 & 0.01 & $-$1.11 & 1.72e$-$07 & 1.74e$-$07 & 11.46 & 11.63 \\
\citep{FRB180301} &  &  & 0.1 & $-$1.17 & 3.86e$-$07 & 3.95e$-$07 & 25.71 & 26.33 \\
 &  &  & 1.0 & $-$1.25 & 6.21e$-$07 & 6.44e$-$07 & 41.39 & 42.96 \\
FRB180212 & 53.0 & 95.9 & 0.01 & $-$1.34 & 8.71e$-$08 & 8.75e$-$08 & 0.09 & 0.09 \\
\citep{ASKAP20} &  &  & 0.1 & $-$1.4 & 2.03e$-$07 & 2.05e$-$07 & 0.21 & 0.21 \\
 &  &  & 1.0 & $-$1.47 & 3.44e$-$07 & 3.48e$-$07 & 0.36 & 0.36 \\
FRB180130 & 23.1 & 94.7 & 0.01 & $-$1.18 & 1.25e$-$06 & 1.35e$-$06 & 1.32 & 1.42 \\
\citep{ASKAP20} &  &  & 0.1 & $-$1.25 & 2.86e$-$06 & 3.14e$-$06 & 3.01 & 3.32 \\
 &  &  & 1.0 & $-$1.32 & 4.79e$-$06 & 5.45e$-$06 & 5.06 & 5.76 \\
FRB180119 & 40.7 & 109.9 & 0.01 & $-$1.31 & 1.07e$-$07 & 1.10e$-$07 & 0.1 & 0.1 \\
\citep{ASKAP20} &  &  & 0.1 & $-$1.38 & 2.36e$-$07 & 2.44e$-$07 & 0.21 & 0.22 \\
 &  &  & 1.0 & $-$1.46 & 3.85e$-$07 & 4.02e$-$07 & 0.35 & 0.37 \\
FRB180110 & 128.1 & 409.9 & 0.01 & $-$1.33 & 2.21e$-$07 & 2.36e$-$07 & 0.05 & 0.06 \\
\citep{ASKAP20} &  &  & 0.1 & $-$1.39 & 5.1e$-$07 & 5.54e$-$07 & 0.12 & 0.14 \\
 &  &  & 1.0 & $-$1.47 & 7.88e$-$07 & 8.74e$-$07 & 0.19 & 0.21 \\
FRB171213 & 88.6 & 132.9 & 0.01 & $-$1.33 & 1.68e$-$07 & 1.77e$-$07 & 0.13 & 0.13 \\
\citep{ASKAP20} &  &  & 0.1 & $-$1.39 & 3.82e$-$07 & 4.08e$-$07 & 0.29 & 0.31 \\
 &  &  & 1.0 & $-$1.47 & 6.14e$-$07 & 6.67e$-$07 & 0.46 & 0.5 \\
FRB171209 & 0.92 & 2.3 & 0.01 & $-$1.28 & 1.9e$-$07 & 1.79e$-$07 & 8.26 & 7.78 \\
\citep{PARKES} &  &  & 0.1 & $-$1.3 & 2.8e$-$07 & 2.98e$-$07 & 12.17 & 12.96 \\
 &  &  & 1.0 & $-$1.32 & 4.5e$$-$$07 & 4.77e$$-$$07 & 19.57 & 20.74 \\
FRB171020 & 117.6 & 376.3 & 0.01 & $-$1.37 & 9e$-$08 & 9.05e$-$08 & 0.02 & 0.02 \\
\citep{ASKAP20} &  &  & 0.1 & $-$1.43 & 1.99e$-$07 & 2.01e$-$07 & 0.05 & 0.05 \\
 &  &  & 1.0 & $-$1.51 & 3.36e$-$07 & 3.4e$-$07 & 0.09 & 0.09 \\
FRB171019 & 40.5 & 218.7 & 0.01 & $-$1.34 & 5.93e$-$08 & 5.92e$-$08 & 0.03 & 0.03 \\
\citep{ASKAP20} &  &  & 0.1 & $-$1.4 & 1.3e$-$07 & 1.3e$-$07 & 0.06 & 0.06 \\
 &  &  & 1.0 & $-$1.48 & 2.22e$-$07 & 2.23e$-$07 & 0.1 & 0.1 \\
FRB171003 & 40.5 & 81.0 & 0.01 & $-$1.25 & 4.42e$-$07 & 4.50e$-$07 & 0.55 & 0.56 \\
\citep{ASKAP20} &  &  & 0.1 & $-$1.32 & 1.01e$-$06 & 1.03e$-$06 & 1.24 & 1.27 \\
 &  &  & 1.0 & $-$1.39 & 1.72e$-$06 & 1.78e$-$06 & 2.13 & 2.2 \\
%FRB170922 & 2.3 & 59.8 & 0.01 & $$$-$$$1.344 & 2.5e$$$-$$$09 & 2.38e$$$-$$$09 & 0.004 & 0.004 \\
% &  &  & 0.1 & $$$-$$$1.33 & 3.5e$$$-$$$08 & 3.58e$$$-$$$08 & 0.059 & 0.06 \\
% &  &  & 1.0 & $$$-$$$1.309 & 5.8e$$$-$$$07 & 5.96e$$$-$$$07 & 0.97 & 0.997 \\
FRB170906 & 29.6 & 74.0 & 0.01 & $-$1.28 & 1.62e$-$07 & 1.68e$-$07 & 0.22 & 0.23 \\
\citep{ASKAP20} &  &  & 0.1 & $-$1.34 & 3.68e$-$07 & 3.86e$-$07 & 0.5 & 0.52 \\
 &  &  & 1.0 & $-$1.42 & 6.01e$-$07 & 6.39e$-$07 & 0.81 & 0.86 \\
FRB170827 & 50.3 & 19.9 & 0.01 & $-$1.32 & 1.10e$-$07 & 1.15e$-$07 & 0.55 & 0.58 \\
\citep{FRB170827} &  &  & 0.1 & $-$1.38 & 2.44e$-$07 & 2.58e$-$07 & 1.23 & 1.3 \\
 &  &  & 1.0 & $-$1.45 & 3.97e$-$07 & 4.25e$-$07 & 2.0 & 2.14 \\
FRB170707 & 14.8 & 51.8 & 0.01 & $-$1.26 & 1.42e$-$07 & 1.46e$-$07 & 0.27 & 0.28 \\
\citep{ASKAP20} &  &  & 0.1 & $-$1.32 & 3.13e$-$07 & 3.25e$-$07 & 0.6 & 0.63 \\
 &  &  & 1.0 & $-$1.4 & 5.05e$-$07 & 5.32e$-$07 & 0.98 & 1.03 \\
FRB170606 & 0.5 & 1782.0 & 0.01 & $-$1.02 & 1.38e$-$06 & 1.39e$-$06 & 0.08 & 0.08 \\
\citet{FRB170606} &  &  & 0.1 & $-$1.08 & 3.13e$-$06 & 3.21e$-$06 & 0.18 & 0.18 \\
 &  &  & 1.0 & $-$1.15 & 4.75e$-$06 & 5.01e$-$06 & 0.27 & 0.28 \\
FRB170428 & 7.7 & 33.9 & 0.01 & $-$1.2 & 2.715e$-$07 & 2.80e$-$07 & 0.8 & 0.83 \\
\citep{ASKAP20} &  &  & 0.1 & $-$1.27 & 6.13e$-$07 & 6.38e$-$07 & 1.81 & 1.88 \\
 &  &  & 1.0 & $-$1.34 & 9.71e$-$07 & 1.03e$-$06 & 2.86 & 3.03 \\
FRB170416 & 19.4 & 97.0 & 0.01 & $-$1.22 & 4.525e$-$07 & 4.79e$-$07 & 0.47 & 0.49 \\
\citep{ASKAP20} &  &  & 0.1 & $-$1.28 & 1.0e$-$06 & 1.085e$-$06 & 1.04 & 1.12 \\
 &  &  & 1.0 & $-$1.36 & 1.56e$-$06 & 1.73e$-$06 & 1.61 & 1.78 \\
FRB160608 & 4.3 & 38.7 & 0.01 & $-$1.11 & 1.88e$-$06 & 1.29e$-$06 & 4.85 & 3.34 \\
\citet{FRB160608} &  &  & 0.1 & $-$1.17 & 4.25e$-$06 & 3e$-$06 & 10.99 & 7.75 \\
 &  &  & 1.0 & $-$1.25 & 6.69e$-$06 & 4.87e$-$06 & 17.29 & 12.57 \\
FRB151230 & 0.4 & 1.9 & 0.01 & $-$1.1 & 1.78e$-$07 & 1.80e$-$07 & 9.36 & 9.49 \\
\citep{FRB151230} &  &  & 0.1 & $-$1.16 & 4e$-$07 & 4.09e$-$07 & 21.05 & 21.54 \\
 &  &  & 1.0 & $-$1.24 & 6.76e$-$07 & 7e$-$07 & 35.57 & 36.87 \\
    \enddata
\end{deluxetable*}

% \appendix

% \section{References and rough notes} \label{app:papers}

% \citet{2016MNRAS.460.2875Y} have put limits on $\xi = [\nu L_\nu]_\gamma / [\nu L_\nu]_\mathrm{radio} \lesssim 4$-$1 \times1-^{7}$. Pulsars have  range of $10^{4}-10^{8}$. Corresponding limit based on \citet{tkp16} for the SGR\,1806$-$20 flare is $\xi \gtrsim 10^{7.5}$. \citet{scholz2017} have put a limit of $3\times 10^{-11}\,\mathrm{erg\,cm^{-2}\,s^{-1}}$ on 0.5--10 keV pulses shorter than 700 ms. From Fermi-LAT data, the 10--100 keV limit is $4\times10^{-9}\,\mathrm{erg\,cm^{-2}\,s^{-1}}$. Burst energy limits $< 10^{45-46}\,\mathrm{erg}$. \citet{tkp16} set limits on FRBs to have ($\eta \gtrsim F_\mathrm{1.4 GHz}/F_{\gamma} > 8 $). \citet{scholz2017} put limits on $\eta > 6 \times 10^9\,\mathrm{Jy\,ms\,erg^{-1}}\,cm^2$ from NuSTAR data and $\eta > 2 \times 10^8\,\mathrm{Jy\,ms\,erg^{-1}\,cm^2}$ from Fermi-LAT data.  
% \citet{xi2017} set limits on Fermi-LAT observations of FRBs till $10^{-6}\,\mathrm{erg\,cm^{-2}}$ however, these are not coincident in time. The time for the FRB location to come into the FoV is $\sim 1000 sec$
% \citet{palaniswamy2014} searched for FRBs post GRBs after delays of ~140 sec.

\bibliographystyle{aasjournal}
\bibliography{references}


\end{document}